\documentclass[floatfix, preprint, showpacs, showkeys, preprintnumbers, nofootinbib, superscriptaddress]{revtex4-1}
\usepackage[utf8]{inputenc}
\usepackage[sort&compress]{natbib}
\usepackage{ulem}
\usepackage{bm}
\usepackage{times}
\usepackage{amssymb,amsbsy,amsmath,amsfonts}
\usepackage{graphicx}
\usepackage{epstopdf}
\usepackage{float}
\usepackage{color}
\usepackage{morefloats}
\usepackage{rotating}
\usepackage{srcltx}
\usepackage{slashed}
\usepackage{subfigure}
\usepackage{multirow}
\usepackage{verbatim}
\usepackage{hyperref}
\usepackage{overpic}
\usepackage{tabularx}

\begin{document}

\title{Can discovery of hidden charm strange pentaquark states help determine the spins of $P_c(4440)$ and $P_c(4457)$?}

\author{Ming-Zhu Liu}
\affiliation{School of Physics, Beihang University, Beijing 100191, China}

\author{Ya-Wen Pan}
\affiliation{School of Physics, Beihang University, Beijing 100191, China}

\author{Li-Sheng Geng}\email{lisheng.geng@buaa.edu.cn}
\affiliation{School of Physics, Beihang University, Beijing 100191, China}
\affiliation{
Beijing Key Laboratory of Advanced Nuclear Materials and Physics,
Beihang University, Beijing 100191, China}
\affiliation{School of Physics and Microelectronics, Zhengzhou University, Zhengzhou, Henan 450001, China}
\affiliation{Beijing Advanced Innovation Center for Big Data-Based Precision Medicine, School of Medicine and Engineering, Beihang University, Beijing, 100191}

\date{\today}
\begin{abstract}
  The pentaquark states, $P_{c}(4312)$, $P_{c}(4440)$ and $P_{c}(4457)$, could
    be nicely arranged into a multiplet of seven molecules of $\bar{D}^{(\ast)}\Sigma_{c}^{(\ast)}$  dictated
     by heavy quark spin symmetry. However, the spins of $P_c(4440)$ and $P_c(4457)$ are not yet fully determined. In this work we employ the contact-range  effective field theory to investigate  the $SU(3)$-flavor counterparts of $\bar{D}^{(\ast)}\Sigma_{c}^{(\ast)}$, and study the possibility whether their discovery can help determine the spins of $P_c(4457)$ and $P_c(4440)$. We find the  existence of a complete hidden charm strange multiplet of $\bar{D}^{(\ast)}\Xi_{c}^{(\prime\ast)}$ molecules irrespective of the spins of $P_c(4440)$ and $P_c(4457)$. On the other hand,  we find that although molecules of $\bar{D}^{(\ast)}\Xi_{c}$    are also likely, depending on the realization of the underlying dynamics, their discovery can be more useful to determine the spins of    $P_{c}(4440)$ and $P_{c}(4457)$ and to  tell how the heavy quark and light quark interaction depends on the spin of the light quark pair.
\end{abstract}


\maketitle
\section{Introduction}

In 2015, the LHCb Collaboration reported two reasonance states, $P_{c}(4380)$ and $P_{c}(4450)$, in the $J/\psi p$ invariant mass spectrum of the $\Lambda_{b}\rightarrow J/\psi p K$ decay~\cite{Aaij:2015tga}, whose mass and decay width are
\begin{eqnarray}
M_{P_{4380}}&=&4380\pm 8\pm 29 ~\mbox{MeV}    \quad \quad   \Gamma_{P_{4380}}=205 \pm 18 \pm 86 ~\mbox{MeV},   \\ \nonumber
M_{P_{4450}}&=&4449.8\pm 1.7\pm 2.5 ~\mbox{MeV}    \quad \quad   \Gamma_{P_{4450}}=39 \pm 5 \pm 19 ~\mbox{MeV},
\end{eqnarray}
respectively. Because of the closeness of $P_{c}(4450)$ to the mass threshold of $\bar{D}^{\ast}\Sigma_{c}$ as well as its narrow decay width,  it is often suggested to be a hadronic molecule~\cite{Chen:2015loa,Roca:2015dva,He:2015cea,Geng:2017hxc}.    It should be noted
  that the existence of $\bar{D}^{(*)}\Sigma_c^{(*)}$ molecules had been predicted before the  LHCb discovery~\cite{Wu:2010jy,Wu:2010vk,Xiao:2013yca,Karliner:2015ina,Wang:2011rga,Yang:2011wz}.  In 2019 the LHCb Collaboration updated their analysis with a data set of  almost ten times bigger and found that the previous $P_{c}(4450)$ state splits into two states, $P_{c}(4440)$ and $P_{c}(4457)$, and in addition a new narrow state $P_{c}(4312)$~\cite{Aaij:2019vzc} emerges just below the $\bar{D}\Sigma_c$ threshold, whose masses and decay widths are
  \begin{eqnarray}
M_{P_{4312}}&=&4311.9\pm 0.7^{+6.8}_{-0.6} ~\mbox{MeV}    \quad \quad   \Gamma_{P_{4312}}=9.8 \pm 2.7^{+3.7}_{-4.5} ~\mbox{MeV},   \\ \nonumber
M_{P_{4440}}&=&4440.3\pm 1.3^{+4.1}_{-4.7} ~\mbox{MeV}    \quad \quad   \Gamma_{P_{4440}}=20.6 \pm 4.9^{+8.7}_{-10.1}~\mbox{MeV},
\\ \nonumber
M_{P_{4457}}&=&4457.3\pm 0.6^{+4.1}_{-1.7} ~\mbox{MeV}    \quad \quad   \Gamma_{P_{4457}}=6.4 \pm 2.0^{+5.7}_{-1.9} ~\mbox{MeV}.
\end{eqnarray}
 In our previous work we showed that these states can be understood as hadronic molecules in both an effective field theory (EFT) approach and the one boson exchange(OBE) model~\cite{Liu:2019tjn,Liu:2019stu}.   Although at present
   the molecular interpretation might be the most favored one~\cite{Xiao:2019aya,Xiao:2019mvs,Sakai:2019qph,Yamaguchi:2019seo,Liu:2019zvb,Valderrama:2019chc,Du:2019pij}, there exist other explanations, e.g.,
hadro-charmonium~\cite{Eides:2019tgv}, compact pentaquark states~\cite{Ali:2019npk,Wang:2019got,Cheng:2019obk,Weng:2019ynv,Zhu:2019iwm,Pimikov:2019dyr}, or
virtual states~\cite{Fernandez-Ramirez:2019koa}. See Refs.~\cite{Liu:2019zoy,Brambilla:2019esw,Guo:2019twa} for some latest reviews. As argued in Ref.~\cite{Pan:2019skd}, the most crucial, but still missing, information to disentangle different interpretations is their spins.

Symmetry is a  core concept in particle physics and plays an important role in studying heavy hadronic molecules. Two  symmetries relevant to the present work are heavy quark spin symmetry (HQSS) and heavy antiquark diquark symmetry(HADS). The HQSS dictates that the strong interaction is independent of the spin of the heavy quark in the limit of heavy quark masses~\cite{Isgur:1989vq,Isgur:1989ed}, which provides a natural explanation to the mass difference of $(D,D^{\ast})$ and $(B,B^{\ast})$, as well as those of their baryon counterparts.  Applying HQSS to studying heavy hadronic molecules, $D_{s0}(2317)$ and $D_{s1}(2460)$ can be naturally interpreted as $DK$ and $D^{\ast}K$ molecules of spin doublets~\cite{Guo:2006rp,Altenbuchinger:2013vwa}. The HQSS also plays an important role in describing  the three pentaquark states, $P_{c}(4312)$, $P_{c}(4440)$, and $P_{c}(4457)$, as $\bar{D}^{\ast}\Sigma_{c}$ molecules. Particularly, we obtained a complete multiplet of hadronic molecules in the $\bar{D}^{(\ast)}\Sigma_{c}^{(\ast)}$ system in  both the EFT approach and OBE model ~\cite{Liu:2019tjn,Liu:2019tjn,Liu:2019stu},  which has been later corroborated by many studies~\cite{Xiao:2019aya,Yamaguchi:2019seo,Liu:2019zvb,Valderrama:2019chc,Du:2019pij}.
In addition to HQSS,  heavy anti-quark di-quark symmetry(HADS) has also been used to study heavy hadronic molecules~\cite{Pan:2019skd,Pan:2020xek} and to estimate the coupling of doubly charmed baryon  to pion~\cite{Liu:2018euh}.  In Refs.~\cite{Pan:2019skd,Pan:2020xek}, we extended the $\bar{D}^{(*)}\Sigma_c^{(*)}$ system to the $\Sigma_c^{(*)}\Xi_{cc}^{(\ast)}$ system via HADS, and predicted a complete multiplet of triply charmed hadronic molecules. In particular, we pointed out that the splitting of $\Xi_{cc}\Sigma_c$ states are correlated with the spins of  $P_{c}(4440)$ and $P_{c}(4457)$, which, given the fact that the former can be much easily simulated on the lattice~\cite{Junnarkar:2019equ}, provides a  possibility to determine the spins of the later in a model independent way.

Along this line, in the present work, we explore whether one can relate the pentaquark states $P_c(4440)$ and $P_c(4457)$ to other states via symmetries such that their discovery could shed light on the nature of the $P_c$ states, particularly their spin in the molecular picture. The symmetry of current interest is  $SU(3)$-flavor symmetry, which relates the hidden charm states to hidden charm strange states. As a result, the discovery of the latter will shed light on the nature of the former.

Before $P_{c}(4380)$ and $P_{c}(4450)$  were discovered by the LHCb collaboration, Wu et al. had already predicted 4 hidden charm strange pentaquark states through local hidden gauge Lagrangian in combination with unitary techniques in couple channels~\cite{Wu:2010jy}. After the discovery of three pentaquark states in 2019, the study was updated and 10 hidden charm strange pentquark states were predicted~\cite{Xiao:2019gjd}.  In Ref.~\cite{Chen:2016ryt} two partners  of $P_{c}(4380)$ and $P_{c}(4450)$ were predicted in the OBE model. More recently, Wang et al. also predicted the existence of 10 hidden charm strange pentaquark states in the chiral effective
field theory~\cite{Wang:2019nvm}.  The discovery potential of hidden charm strange pentaquark states in the $J/\psi \Lambda$ invariant mass spectra of the  $\Xi_{b}\rightarrow J/\psi \Lambda K$ and $\Lambda_{b}\rightarrow J/\psi \Lambda K$ decays have been explored~\cite{Lu:2016roh,Chen:2015sxa}, as well as their partial decay widths~\cite{Shen:2019evi}.

In this work we  adopt an effective field theory (EFT) approach to study possible hidden charm  strange molecules of $\bar{D}^{(\ast)}\Xi_{c}$ and  $\bar{D}^{(\ast)}\Xi_{c}^{\prime\ast}$, which can be regarded as the $SU(3)$-flavor counterparts of $\bar{D}^{(\ast)}\Sigma_{c}^{(\ast)}$. In particular, we focus on the correlation between hidden charm and hidden charm strange molecules, which not only can help check the molecular interpretation of $P_{c}(4440)$ and $P_{c}(4457)$  but also can help determine their spin ordering if the hidden charm strange pentaquark states are discovered by either experiments or lattice QCD calculations, analogous to the correlation dictated by HADS as shown in Refs.~\cite{Pan:2019skd,Pan:2020xek}.

  The manuscript is structured as follows. In Sec.~\ref{sec:obe} we present the details of the contact-range potential of $\bar{D}^{(\ast)}\Xi_{c}$ and  $\bar{D}^{(\ast)}\Xi_{c}^{\prime\ast}$ according to heavy quark spin symmetry and SU(3) flavor symmetry.
In Sec.~\ref{sec:pre}
we give the full spectrum of hidden charm strange  molecules.
Finally we present the conclusions in Sec.~\ref{sum}

\section{Theoretical formalism}
\label{sec:obe}
Here we explain how to determine the  $\bar{D}^{(\ast)}\Xi_{c}$ and $\bar{D}^{(\ast)}\Xi_{c}^{\prime\ast}$ interactions and study the likely existence of  hidden charm strange pentaquark states. In the line of Refs. \cite{Liu:2019tjn,Pan:2019skd}, their interactions can be determined in an EFT approach. One should note that we just consider the leading order contact-range potentials because our previous studies indicated that the  pion exchange contributions
are perturbative in the charm sector~\cite{Valderrama:2012jv,Lu:2017dvm}.

Because the $\Xi_{c}^{\prime\ast}$ and $\Sigma_{c}^{(\ast)}$ baryons  belong to the same $SU(3)$ group representation, the interactions of $\bar{D}^{(\ast)}\Xi_{c}^{\prime\ast}$ and $\bar{D}^{(\ast)}\Sigma_{c}^{(\ast)}$ are the same in the heavy quark mass and $SU(3)$-flavor symmetry limits.  As a result, the same two low energy constants are needed to account for the contact-range  potential of $\bar{D}^{(*)}\Sigma_c^{(*)}$ and $\bar{D}^{(\ast)}\Xi_{c}^{\prime(\ast)}$, $C_{a}$ and $C_{b}$, namely,
\begin{equation}
V=C_a + \sigma_1\cdot\sigma_2 C_b.
\end{equation}
There are seven combinations for the $\bar{D}^{(\ast)}\Xi_{c}^{\prime(\ast)}$ system according to HQSS, whose potential can be written as
\begin{eqnarray}
  V(\bar{D} \Xi_c^{\prime}, J = \tfrac{1}{2}) &=& C_a \, , \\
  \label{eq:contact-DSigma-1}
  \nonumber \\
  V(\bar{D} \Xi_c^*, J = \tfrac{3}{2}) &=& C_a \, , \\
  \nonumber \\
  V(\bar{D}^* \Xi_c^{\prime}, J = \tfrac{1}{2}) &=& C_a - \tfrac{4}{3}\,C_b \, , \\
  V(\bar{D}^* \Xi_c^{\prime}, J = \tfrac{3}{2}) &=& C_a + \tfrac{2}{3}\,C_b \, , \\
  \nonumber \\
  V(\bar{D}^* \Xi_c^*, J = \tfrac{1}{2}) &=& C_a - \tfrac{5}{3}\,C_b \, , \\
  V(\bar{D}^* \Xi_c^*, J = \tfrac{3}{2}) &=& C_a - \tfrac{2}{3}\,C_b \, , \\
  V(\bar{D}^* \Xi_c^*, J = \tfrac{5}{2}) &=& C_a + C_b \, .
  \label{eq:contact-DSigma-2}
\end{eqnarray}

The interaction of the $\bar{D}^{(\ast)}\Xi_{c}$ system is different from that of   $\bar{D}^{(\ast)}\Xi_{c}^{\prime\ast}$  because the spin of the light quark pair in $\Xi_{c}$ is 0 and that in $\Xi_{c}^{\prime\ast}$ is 1.
In terms of HQSS, the contact-range potential between $\bar{D}^{(\ast)}$ and  $\Xi_{c}^{\prime\ast}$ can be denoted as $F_{1/2}$  and $F_{3/2}$  via the  coupling of the light quark spins, i.e.,  $1/2 \otimes 1 = 1/2 \oplus 3/2$. Applying the same approach to $\bar{D}^{(\ast)}\Xi_{c}$, the corresponding potential can be represented by one low energy constant $F_{1/2}^{\prime}$ which is from the light quark spin coupling, $1/2 \otimes 0 = 1/2 $.

In principle, $F_{1/2}$ and $F_{1/2}^\prime$ can be different, and there is no reliable way to relate them. In this work, we will take two assumptions and rely on future experiments or lattice QCD simulations to verify which assumption is realized in nature.

First, we assume that  $F_{1/2}$ in the $\bar{D}^{(\ast)}\Xi_{c}^{\prime\ast}$ system  and $F_{1/2}^{\prime}$ in the $\bar{D}^{\ast}\Xi_{c}$ system are the same, which is denoted as Case I in the following.   To find the relationship between the $\bar{D}^{(\ast)}\Xi_{c}^{\prime\ast}$  system and the $\bar{D}^{(\ast)}\Xi_{c}$ system,   the couplings of $F_{1/2}$  and $F_{3/2}$ can  be rewritten as
$F_{1/2}=C_{a}-2C_{b}$ and  $F_{3/2}=C_{a}+C_{b} $.
Then the contact-range potentials of $\bar{D}^{(\ast)}\Xi_{c}$ have the following form
\begin{eqnarray}
  V(\bar{D} \Xi_c, J = \tfrac{1}{2}) &=& C_a - 2\,C_b \, , \\
  \label{eq:contact-DSigma-1}
  \nonumber \\
  V(\bar{D}^{\ast} \Xi_c, J = \tfrac{1}{2}) &=& C_a - 2\,C_b \, , \\
  V(\bar{D}^{\ast} \Xi_c, J = \tfrac{3}{2}) &=& C_a - 2\,C_b \, .
  \label{eq:contact-DSigma-2}
\end{eqnarray}
One should note that the $\bar{D}^{(\ast)}\Xi_{c}^{\prime(\ast)}$ contact-range potential can also be denoted as $F_{1/2}$ and $F_{3/2}$ in terms of HQSS.

 Second, we assume that $F_{1/2}^\prime$  is not the same as $F_{1/2}$, and turn to some phenomenological methods for help, which is denoted as Case II. One of such phenomenological methods is the local hidden gauge approach. According to Ref.~\cite{Xiao:2019gjd}, the contact-range potential of the $\bar{D}^{(\ast)}\Xi_{c}$ system is written as
\begin{eqnarray}
  V(\bar{D} \Xi_c, J = \tfrac{1}{2}) &=& C_a  , \\
  \label{eq:contact-DSigma-1}
  \nonumber \\
  V(\bar{D}^{\ast} \Xi_c, J = \tfrac{1}{2}) &=& C_a , \\
  V(\bar{D}^{\ast} \Xi_c, J = \tfrac{3}{2}) &=& C_a  .
  \label{eq:contact-DSigma-2}
\end{eqnarray}
Clearly, the strength is the same for all the three channels, but it is different from that of Case I. We hope that future experimental or lattice QCD data will tell which case is realized in nature.

To search for bound states, we solve the Lippmann-Schinwinger
equation with contact range potentials
\begin{eqnarray}
\phi(k)+\int \frac{d^{3}p}{(2\pi)^3}\langle
k|V|p\rangle\frac{\phi(p)}{B+\frac{p^2}{2\mu}}=0, \label{10}
\end{eqnarray}
where $\phi(k)$ is the vertex function, $B$ the binding energy, and $\mu$
the reduced mass. To solve the equation we have to regularize the
contact potential
\begin{eqnarray}
\langle p| V_{\Lambda}|
p^{\prime}\rangle=C(\Lambda)f(\frac{p}{\Lambda})f(\frac{p^{\prime}}{\Lambda}),
\end{eqnarray}
with $\Lambda$ the cutoff, $f(x)$ a regular function, and $C({\Lambda})$ the
running coupling constant.  A typical choice of the cutoff is
$\Lambda=0.5-1$ GeV, while for the regulator we choose  a gaussian type
$f(x)=e^{-x^2}$. In this work, we only consider $S$-wave
contact contribution, thus the integral equation simplifies
to
\begin{eqnarray}
1+C(\Lambda)\frac{\mu}{\pi^2}\int_{0}^{\infty}dqe^{-2\frac{q^2}{\Lambda^{2}}}\frac{q^2}{B+\frac{\vec{q}^{2}}{2\mu}}=0.
\label{10}
\end{eqnarray}

\section{Numerical results and discussions}
\label{sec:pre}

From HQSS and SU(3)-flavor symmetry, we express the contact potentials of
$\bar{D}^{(\ast)}$ and $\Xi_{c}^{\prime \ast}$ by two coupling constants,
$C_{a}$ and $C_{b}$, which are the only two unknown inputs for us to
obtain the binding energies of meson-baryon systems under consideration. In our previous
works,  we proposed that $P_{c}(4440)$ and $P_{c}(4457)$
are bound states  of $\bar{D}^{\ast}\Sigma_{c}$  with
either spin 1/2 or 3/2 and negative party, and the two LECs $C_a$ and $C_b$ have been determined by fitting to the masses of $P_c(4440)$ and $P_c(4457)$. As a result, in the following, we study two scenarios, Scenario A where  $P_{c}(4440)$ and
$P_{c}(4457)$ have $1/2^{-}$ and $3/2^{-}$, and Scenario B the other way around.  The results are displayed in Table~\ref{resutls}. We find that all the seven states in the $\bar{D}^{(\ast)}\Xi_{c}^{\prime \ast}$ system bind in both scenario A and B, consistent with  Refs. \cite{Xiao:2019gjd} and \cite{Wang:2019nvm}. It indicates that such kinds of hidden charm strange molecules must exist. In addition   their results favor our scenario A, namely, $P_{c}(4440)$ and $P_{c}(4457)$ have spin 1/2 and 3/2, respectively.

From Table~\ref{resutls} one can easily see that our results are almost independent of the cutoff. Compared to the results of the $\Xi_{cc}^{(\ast)}\Sigma_{c}^{(\ast)}$ bound states~\cite{Pan:2019skd}, the cutoff dependence is much weaker, which implies that $SU(3)$-flavor symmetry  is less broken than HADS.
To estimate the breaking of $SU(3)$-flavor symmetry, we supplement the potentials with a $15\%$  uncertainty. The results taking into account the breaking show that all the states  still bind and the spin orderings remain unchanged, which suggests that the hidden charm strange molecules must exist if $P_{c}(4312)$, $P_{c}(4440)$ and $P_{c}(4457)$ are (dominantly) $\bar{D}^{(\ast)}\Sigma_{c}$ molecules.
   Compared with their hidden charm partners, these states could be detected in the $J/\psi \Lambda$ invariant mass spectra of the
$\Xi_{b}\rightarrow J/\psi \Lambda K$ decay.  If these states are discovered experimentally, it will not only further
enrich the family of hadronic molecules but also helps to determine the spin ordering of $P_{c}(4440)$ and $P_{c}(4457)$.

   \begin{table}[!h]
\caption{Bound states of a singly charmed baryon and a singly charmed antimeson, obtained in the contact range effective field theory with the two constants fixed by reproducing  $P_{c}(4440)$ and $P_{c}(4457)$ as $1/2^-$ and $3/2^-$ molecules (Scenario A) and $3/2^-$ and $1/2^-$ (Scenario B), respectively,
with cutoffs of 0.5 and 1 GeV  \label{resutls} }
\centering
\begin{tabular}{c|cc|cc|cc|c|ccccc}
\hline\hline State& $J^{P}$  & $\Lambda$(GeV)   & B. E(A) & Mass(A) &  B. E(B) & Mass(B) & Ref.~\cite{Xiao:2019gjd}  &  Ref.~\cite{Wang:2019nvm}   \\
\hline $\bar{D}\Xi_{c}^{\prime}$ &$1/2^{-}$ &
1(0.5) & $8.5^{+9.7}_{-6.5}(9.3_{-4.4}^{+5.0})$  &4437(4436)  & $14.0^{+12.2}_{-9.1}(14.9_{-5.9}^{+6.7})$    & 4431(4430)    & 4436.7  &  4423.7
\\
  \hline
$\bar{D}\Xi_{c}^{\ast}$ &$3/2^{-}$ & 1(0.5)&
$9.0^{+9.9}_{-6.7}(9.5_{-4.4}^{+5.1})$ &4504(4504)   & $14.7^{+12.4}_{-9.4}$($15.2_{-5.9}^{+6.7}$)   &   4499(4498)     & 4506.99  & 4502.9
\\ \hline
$\bar{D}^{\ast}\Xi_{c}^{\prime}$ &$1/2^{-}$ &1(0.5)  & $23.4^{+15.5}_{-12.6}(22.5_{-7.7}^{+8.3})$     &4563(4564)  &   $5.6^{+7.9}_{-4.7}$$(5.2_{-3.0}^{+3.7})$  &4581(4581)  & 4580.96  &4568.7
\\
$\bar{D}^{\ast}\Xi_{c}^{\prime}$& $3/2^{-}$
 &1(0.5) & $5.6^{+7.9}_{-4.7}(5.2_{-2.9}^{+3.7})$ & 4581(4581)  & $23.4^{+15.5}_{-12.6}$($22.5_{-7.7}^{+8.3}$)  &  4563(4564)
 &  4580.96  &  4582.3
\\
\hline $\bar{D}^{\ast}\Xi_{c}^{\ast}$ &$1/2^{-}$
&1(0.5)& $28.0^{+16.9}_{-15.1}(26.3_{-8.5}^{+9.1})$ & 4627(4628)  & $4.0^{+6.7}_{-3.7}$ $(3.3_{-2.1}^{+2.9})$  & 4651(4651)  & 4650.86  & 4635.4
\\
 $\bar{D}^{\ast}\Xi_{c}^{\ast}$ &$3/2^{-}$
& 1(0.5)& $17.2^{+13.1}_{-10.2}(16.4_{-6.2}^{+6.8})$ &4637(4638)  & $11.1^{+10.6}_{-7.6}$$(10.5_{-4.6}^{+5.2})$   &  4643(4644)  & 4650.58 & 4644.4
\\
 $\bar{D}^{\ast}\Xi_{c}^{\ast}$ &$5/2^{-}$
& 1(0.5)& $4.0^{+6.7}_{-3.7}(3.3_{-2.1}^{+2.9})$ & 4651(4651)  & $28.0^{+16.9}_{-14.1}$$(26.3_{-8.5}^{+9.1})$  & 4627(4628)  & 4650.56  & 4651.7
\\
\hline\hline
\end{tabular}
\end{table}

For the $\bar{D}^{(\ast)}\Xi_{c}$ system, Case I
assumes that the coupling $F_{1/2}$  is the same as the coupling $F_{1/2}^{\prime}$, and therefore the contact-range potential of $\bar{D}^{(\ast)}\Xi_{c}$ can also be written as combinations of $C_{a}$ and $C_{b}$.    Thus we can calculate the binding energies of the $\bar{D}^{(\ast)}\Xi_{c}$ systems in the two scenarios A and B as well. The results are shown in Table~\ref{resutls5}.  Interestingly, we note that the binding energies  in Scenario A are much larger than their counterparts in Scenario B.  To estimate the uncertainty of $SU(3)$-flavor symmetry and the assumed equality of $F_{1/2}$  and $F_{1/2}^{\prime}$, we consider a larger uncertainty of $30\%$ into the $C_a$ and $C_b$ values.  We find that that the $\bar{D}^{(\ast)}\Xi_{c}$ systems in Scenario B can become unbound, while they still bind in Scenario A, which indicates that the $\bar{D}^{(\ast)}\Xi_{c}$ hidden charm strange molecules exist in Scenario A and may not necessarily exist in Scenario B.   If such molecules are found experimentally, it implies that the spins of  $P_{c}(4440)$ and $P_{c}(4457)$  are more likely to be 1/2 and 3/2, respectively.

In  Case II we estimate the coupling $F_{1/2}'$ by turning to the local hidden-gauge approach. As shown above, the value is different from that of Case I. The corresponding results  are tabulated in Table \ref{resutls5}. We find that the differences between the binding energies in scenario A and those in Scenario B become less extreme, which implies that we can not discriminate the spins of $P_{c}(4440)$ and $P_{c}(4457)$ in Case II.
As a result, it may not help much to derive the spin ordering of $P_c(4457)$ and $P_c(4440)$ even these states are discovered experimentally. On the other hand, their discovery does help to reveal more the nature of the hidden charm pentaquark states.

\begin{table}[!h]
\caption{(Likely) bound states of a singly charmed baryon and a singly charmed antimeson, obtained in the contact range effective field theory with the two constants fixed either by reproducing $P_{c}(4440)$ and $P_{c}(4457)$ as $1/2^-$ and $3/2^-$ molecules (Scenario A) and $3/2^-$ and $1/2^-$ (Scenario B), respectively.
with cutoffs of 0.5 and 1 GeV. Case I and II differ from each other in that how $F'_{1/2}$ is determined.  \label{resutls5} }
\centering
\begin{tabular}{c|c|cc|cc|cc|c|ccccc}
\hline\hline  &  state& $J^{P}$  & $\Lambda$(GeV)   & B. E(A) & Mass(A) &  B. E(B) & Mass(B) & \cite{Xiao:2019gjd}  &  \cite{Wang:2019nvm}   \\
\hline
\multirow{3}{0.8cm}{I}&   $\bar{D}\Xi_{c}$ &$1/2^{-}$ &
1(0.5) &  $26.3_{-24.3}^{+36.1}(27.4^{+19.6}_{-16.9}) $   & 4310(4309)  & $0.9^{+10.5}_{\dag}(1.0^{+4.1}_{\dag})$    & 4335(4335)    & 4276.59 & 4319.4
\\
 &  $\bar{D}^{\ast}\Xi_{c}$ &$1/2^{-}$ &1(0.5)  & $29.5_{-25.4}^{+37.4}(28.8^{+20.0}_{-17.4})$     &4448(4449)  &   $1.6^{+12.0}_{\dag}(1.3^{+4.5}_{\dag})$  &4476(4476)  & 4429.84  &4456.9
\\
 &    $\bar{D}^{\ast}\Xi_{c}$& $3/2^{-}$
 &1(0.5)  & $29.5_{-25.4}^{+37.4}(28.8^{+20.0}_{-17.4})$     &4448(4449)  &   $1.6^{+12.0}_{\dag}(1.3^{+4.5}_{\dag})$  &4476(4476)  & 4429.84  &4456.9  \\ \hline
\multirow{3}{0.8cm}{II} &  $\bar{D}\Xi_{c}$ &$1/2^{-}$ &
1(0.5) &  $7.7_{\dag}^{+20.9}(8.9^{+10.5}_{-7.4})$   & 4329(4327)  & $13.0^{+26.0}_{-12.9}$($14.4_{-10.6}^{+13.6}$)    & 4335(4321)    & 4276.59 & 4319.4
\\
&  $\bar{D}^{\ast}\Xi_{c}$ &$1/2^{-}$ &1(0.5)  & $9.6_{\dag}^{+22.4}(9.8^{+10.8}_{-7.9})$     &4468(4468)  &   $15.4^{+28.4}_{-15.0}$($15.5_{-11.0}^{+14.0}$)  &4462(4462)  & 4429.84  &4456.9
\\
 & $\bar{D}^{\ast}\Xi_{c}$& $3/2^{-}$
 &1(0.5) & $9.6_{\dag}^{+22.4}(9.8^{+10.8}_{-7.9})$     &4468(4468)  &   $15.4^{+28.4}_{-15.0}$($15.5_{-11.0}^{+14.0}$)  &4462(4462)  & 4429.84  &4456.9
  \\
\hline\hline
\end{tabular}
\end{table}

\section{Summary}
\label{sum}
In this work assuming that $P_{c}(4312)$, $P_{c}(4440)$ and  $P_{c}(4457)$ are $\bar{D}^{(\ast)}\Sigma_{c}^{(\ast)}$ molecules, we employed a contact-range effective field theory approach satisfying HQSS and SU(3)-flavor symmetry to study the likely existence of hidden charm strange molecules with the main purpose to determine the spin ordering of $P_c(4440)$ and $P_c(4457)$. To estimate the uncertainty caused by the breaking of these symmetries, we considered a $15\%$ breaking for $\bar{D}^{(\ast)}\Xi_{c}^{\prime\ast}$  and a $30\%$ breaking for $\bar{D}^{(\ast)}\Xi_{c}$ in our study.
Our results showed that there exists a complete multiplet of hadronic molecules in the $\bar{D}^{(\ast)}\Xi_{c}^{\prime(\ast)}$ system irrespective of the spins of $P_c(4440)$ and $P_c(4457)$.

Assuming that the couplings  $F_{1/2}$ and $F_{1/2}^{\prime}$   are the same, the existence of
$\bar{D}^{(\ast)}\Xi_{c}$ molecules is only likely if $P_c(4457)$ has spin 3/2 and $P_c(4440)$ has spin 1/2. As a result, the discovery of these states can help to determine the spins of
$P_{c}(4440)$ and  $P_{c}(4457)$ in the molecular picture from an EFT perspecive. On the other hand, using the hidden gauge approach to infer the coupling of $F_{1/2}'$,  $\bar{D}^{(\ast)}\Xi_{c}$ molecules exist irrespective of the spins of $P_{c}(4440)$ and  $P_{c}(4457)$, which thus offers little help in determining their spins.

Note added in proof: In a recent talk, the LHCb Collaboration has reported the likely existence of a hidden charm strange pentaquark $P_{cs}(4459)$ with a statistical significance of 3.1 sigma~\cite{Pan:2020123456}, which has inspired a number of theoretical studies~\cite{Chen:2020uif,Wang:2020eep,Peng:2020hql}.

\section{Acknowledgments}
 This work is partly supported by the National Natural Science Foundation of China under Grants Nos.11735003, 11975041, and 11961141004, and the fundamental Research Funds
for the Central Universities.

\bibliography{XiccSigmac}

\end{document}